\documentstyle[12pt]{article}
\begin{document}
\baselineskip=24pt
\newcommand{\kq}{\frac{k}{q}}
\newcommand{\kqp}{\frac{k'}{q'}}
\newcommand{\di}{\displaystyle}
\newcommand{\ekq}{e^{2\pi i\frac{k}{q}}}
\newcommand{\rf}{Ramanujan - Fourier~}
\newcommand{\ekqn}{e^{2\pi i\frac{k}{q}n}}
\newcommand{\ekqm}{e^{-2\pi i\frac{k}{q}}}
\newcommand{\ekqmn}{e^{-2\pi i\frac{k}{q}n}}
\newcommand{\wk}{Wiener - Khintchine formula~}
\newcommand{\ekqp}{e^{2\pi i\frac{k'}{q'}}}
\begin{center}
{\Large \bf Renormalisation and the density \\
of prime pairs}\\
\vspace{1cm}
{\bf G.H. Gadiyar \footnote {E-mail: padma@imsc.ernet.in} and R. Padma $^{(a)}$ \footnote {E-mail: padma@imsc.ernet.in}}\\
{\it $^{(a)}$ Ramanujan Institute for Advanced Study in Mathematics,\\
University of Madras, Chennai 600 005 INDIA}
\end{center}

\vspace{3cm}

\noindent {\bf Abstract}

Ideas from physics are used to show that the prime pairs have the density 
conjectured
by Hardy and Littlewood. The proof involves dealing with infinities like in
quantum field theory. 

\vspace{1cm}


\newpage
This article may be considered as an invitation to theoretical physicists
to enter the field of additive number theory.
For sometime now there have been serious attempts to cross fertilise the
disciplines of physics and number theory. It seems strange that on
the one hand 
the most practical of disciplines, namely, physics has connections with
the most aracane of disciplines, namely, number theory. However,
surprising connections have appeared between number theory and physics
as can be seen in [1], [3] and [9]. The work of Ramanujan in particular has
had surprising connections with string theory, conformal field theory
and statistical physics.

For sometime now the authors one of whom is a theoretical physicist and the
other a number theorist have been trying to understand problems in additive
number theory using ideas from both fields. One such problem is the distribution
of prime pairs. Prime pairs are numbers which are primes differing by some even
integer. For example, $3,5; 5,7; 11, 13; 17, 19$ and so on are all prime pairs
with common difference $2$. The question is whether such prime pairs are 
infinite, if so, what is the density?

We will now summarise the standard method used to attack the problem which is the
circle method. For technical reasons, the von Mangoldt function $\Lambda (n)$ (which is defined to be $\log p$ if $n=p^m$ where $p$ is a prime and $0$ otherwise) is
used instead of the characteristic function on the primes. Hence in the circle
method
$$
\int_0^1 \vert \sum_{n=1}^{\infty} \Lambda (n) e^{2\pi in\theta} \vert ^2
e^{-2\pi ihz} dz 
$$
gives the number of prime pairs with common difference $h$, an even integer.
The interval of integration $[0,1)$ in the above expression is partitioned
by certain rational numbers called Farey fractions. The circle method
consists in proving that the  sum of the integrals over the intervals that
correspond to the fractions with small denominators (called major arc) gives
the conjectural main term and the contribution from the remaining part of
the interval $[0,1)$ (called minor arc) is small. 

The authors noted in an earlier preprint [4] that the entire method could be
replaced by proving a Wiener - Khintchine formula for arithmetical functions
having Ramanujan - Fourier expansion. 
Ramanujan [7] using extremely simple
arguments showed that the commonly known arithmetical functions all have
\rf ~expansions. These are typically expansions of the form
$$
a(n) ~=~ \sum_{q=1}^{\infty} a_q c_q(n)  ~ ,  \eqno (1)
$$
where 
$$
c_q(n) ~=~ \sum_{\stackrel{k=1}{(k,q)=1}}^{q}  \ekqn \, .
$$
and is known as the Ramanujan sum.
For example he showed that 
\begin{eqnarray*}
d(n) &=& - \sum_{q=1}^{\infty} \frac{\log q}{q} c_q(n) ~,\\
\sigma (n) &=& \frac{\pi ^2}{6} n \sum_{q=1}^{\infty} \frac{c_q(n)}{q^2} ~,\\
r(n) &=& \pi \sum_{q=1}^{\infty} \frac{(-1)^{q-1}}{2q -1} c_{2q-1}(n) 
\end{eqnarray*}
where $d(n)$ is the number of divisors of $n$, $\sigma (n)$ their sum and
$r(n)$ the number of ways $n$ can be expressed as the sum of two squares.
Hardy in an extremely elegant paper [5] elucidated the arithmetical properties of $c_q(n)$.
He proved that $c_q(n)$ is multiplicative. That is,
$$
c_{qq'} (n) = c_q (n) c_{q'} (n) \, , ~{\rm if}~ (q,q')=1
$$
and using this property he proved that
$$
\frac{\phi (n)}{n} \Lambda (n) = \sum_{q=1}^{\infty} \frac{\mu (q)}{\phi (q)} c_q (n)
\eqno (2) 
$$
where $\mu (q)$ is the M\"{o}bius function defined as follows: 
$\mu (1) =1,~ \mu (n)=0,$ if $n$ has a squared factor and
$\mu (p_1 p_2....p_l) ~=~ (-1)^l$ if $p_1,~p_2,.....p_l$ are different primes and $\phi (q)$ is the Euler -
totient function defined as the number of positive integers less than
and prime to $n$.  
The next step was taken by Carmichael [2] who essentially showed that
$\ekqn$ are almost periodic functions defined on the integers. This
leads to orthogonality relations and a method for evaluating the
\rf coefficient. Denote by $M(g)$ the mean value of $g$, that is,
$$
M(g) ~=~ \lim_{N \rightarrow \infty} \frac{1}{N} \sum_{n \le N} g(n) \, .
$$
For $1 \le k \le q, ~ (k,q)=1,$ let
$e_{\frac{k}{q}}(n)=e^{2\pi i \frac{k}{q} n}$, ($n \in \cal N$). If
$a(n)$ is an arithmetical function with the Ramanujan - Fourier expansion (1)
then
$$
a_q ~=~ \frac{1}{\phi (q)} M(a c_q) ~= \frac{1}{\phi (q)}\lim_{N \rightarrow \infty} \frac{1}{N}
\sum_{n \le N} a(n) c_q(n)  \, . 
$$
Also 
$$
M(e_{\frac{k}{q}}  ~\overline{e_{\frac{k'}{q'}}}) = \left \{ \begin{array}{ll}
1 \, , &if~~ \frac{k}{q}~=~ \frac{k'}{q'} \, ,\\
0 \, , & if~~ \frac{k}{q} ~\ne ~\frac{k'}{q'} \, .
\end{array} \right . \eqno (3)
$$
All this is well known in the field of almost periodic functions.

Hardy and Littlewood in [6] conjectured that there are infinitely many prime pairs $p,
p+h$ for every even integer $h$ and if $P_h(N)$ denotes the number of pairs
less than $N$, then 
$$
P_h(N) \sim 2 \prod_{p>2} \left (1-\frac{1}{(p-1)^2}\right ) \frac{N}{\log^2N}
\prod_{\stackrel{p | h}{p > 2}} \left (\frac{p-1}{p-2}\right ) \, .
$$
This is equivalent to 
$$
\sum_{n \le N} \Lambda(n) \Lambda(n+h)  
\sim 2 N \prod_{p>2} \left (1-\frac{1}{(p-1)^2}\right )
\prod_{\stackrel{p | h}{p >2}} \left (\frac{p-1}{p-2}\right ) \, .
$$
Numerical calculations done by Mrs. Streatfeild [6] strengthen the validity
of this conjecture.

Using (2) and (3) one can give the heuristic argument
\begin{eqnarray*}
\lim_{N \rightarrow \infty}&{\displaystyle \frac{1}{N}}& \sum_{n\le N} \frac{\phi(n)}{n}\Lambda (n) \frac{\phi(n+h)}{n+h} \Lambda (n+h) \\
&=& \lim_{N \rightarrow \infty} \frac{1}{N} \sum_{n \le N} 
\sum_{q=1}^{\infty} \sum_{\stackrel{k=1}{(k,q)=1}}^q 
\sum_{q'=1}^{\infty} \sum_{\stackrel{k'=1}{(k',q')=1}}^{q'} \\
& & \left ( \frac{\mu(q)}{\phi(q)} e^{2\pi i \frac{k}{q}n}
\frac{\mu(q')}{\phi(q')} e^{-2\pi i \frac{k'}{q'}(n+h)} \right )\\
&=&\sum_{q=1}^{\infty} \sum_{\stackrel{k=1}{(k,q)=1}}^q 
\sum_{q'=1}^{\infty} \sum_{\stackrel{k'=1}{(k',q')=1}}^{q'} 
\frac{\mu(q)} {\phi(q)} 
\frac{\mu(q')}{\phi(q')} e^{-2\pi i \frac{k'}{q'}h}
M(e_{\kq}, \overline{e_{\kqp})}\\
&=& \sum_{q=1}^{\infty} \frac{\mu ^2(q)}{\phi ^2 (q)} c_q(h) ~~ , \\
&=& 2\prod_{p>2} \left (1-\frac{1}{(p-1)^2} \right ) \prod_{\stackrel{p|h}{p>2}}\left (\frac{p-1}{p-2}\right )  ~~ .
\end{eqnarray*}

However the proof of the Wiener -
Khintchine formula is not possible for the von Mangoldt function due to the
poor estimates available for  ${\di \frac{1}{N} \sum_{n \le N}}  e^{\di {2\pi i  
(\kq - \kqp)n}}$ where $1 \le k \le q$ and $(k,q) =1$ and
$1 \le k' \le q'$ and $(k',q') =1$ and $\kq \neq \kqp$.

In this paper we show that replacing ${\di \sum_{n=0}^{\infty}} a_n e^{in\theta} $ by
${\di \sum_{n= - \infty}^{\infty}} a_n e^{in\theta} $ (note that the limits of summations
are different) makes it unnecessary to have to undertake error term analysis.

Consider
\begin{eqnarray*}
~~ && \sum_{N=1}^{\infty} \sum_{n \le N} \frac{\phi(n)}{n} \Lambda (n)
\frac{\phi(n+h)}{n+h} \Lambda (n+h) x^N\\
~ &=& \sum_{N=1}^{\infty} \sum_{n \le N} \sum_{q,q'=1}^\infty \frac{\mu(q)}{\phi(q)}
\frac{\mu (q')}{\phi (q')}c_q(n) c_{q'}(n+h) x^N\\
~ &=& \sum_{q,q'=1}^{\infty} \sum_{\stackrel{k=1}{(k,q)=1}}^q\sum_{\stackrel{k'=1}{(k',q')=1}}^{q'}
\sum_{n=1}^{\infty}\frac{\mu(q)}{\phi(q)}\frac{\mu(q')}{\phi(q')} 
e^{2\pi i ((\kq - \kqp)n -\kqp h)} \sum_{N=n}^{\infty} x^N\\
~ &=& \frac{1}{1-x}\sum_{q,q'=1}^{\infty} \sum_{\stackrel{k=1}{(k,q)=1}}^q\sum_{\stackrel{k'=1}{(k',q')=1}}^{q'}
\sum_{n=1}^{\infty} \frac{\mu(q)}{\phi(q)}\frac{\mu(q')}{\phi(q')} 
e^{2\pi i ((\kq - \kqp)n -\kqp h)} x^n \\
~ &=& \frac{x}{(1-x)^2} \sum_{q=1}^{\infty} \frac{\mu ^2 (q)}{\phi ^2 (q)} c_q(h)
~+~ {\rm cross~ terms}  \, . \hspace{4cm} (4)
\end{eqnarray*}
Thus we have,
$$
\sum_{n \le N} \frac{\phi(n)}{n} \Lambda (n)
\frac{\phi(n+h)}{n+h} \Lambda (n+h) =  N~ \sum_{q=1}^{\infty} \frac{\mu ^2 (q)}{\phi ^2 (q)} c_q(h) ~+~ {\rm cross ~ terms} ~ .
$$
Note that the first term is the main term as conjectured by Hardy and Littlewood. However,
the toughest part is to show that the main term which are few in number is
greater than the error terms which are greater in number. For this even assuming
hypotheses like those of Riemann or Montgomery and Vaughan do not seem to suffice. {\it {Analytical number theorists spend most of their time and effort
in estimates of these error terms. }}

In this paper we just make a small change in the ideas of Ramanujan, Hardy and
Littlewood. We note first the Poisson summation formula in the distributional
form [8].
$$
\sum_{n= - \infty}^{\infty} e^{in \theta} =\sum_{k = - \infty}^{\infty}  
\delta ( \theta - 2\pi k)
$$
It is immediately obvious that in (4), summation from 1 to $\infty $ gives
the cross terms which contribute to 
error term. {\it {But in this case because one has to integrate product of $ \delta $
functions, there are no cross terms. However this leads to a new
difficulty which seems tractable using ideas from physics like
renormalisation. }}

Let $a(n)$ be an arithmetical function having the Ramanujan - Fourier expansion (1). Since $c_q(n)~=~c_q(-n)$, we take $a(-n)~=~a(n)$. Then 
\newpage
\begin{eqnarray*}
~ &&\sum_{n= -\infty}^{\infty} a_n a_{n+h} \\
&=& \frac{1}{2\pi} \int_0^{2 \pi}  \left ( \sum_{n= -\infty}^{\infty} a_n e^{in\theta}  \right )~ 
\left ( \sum_{m= -\infty}^{\infty} a_m e^{-im\theta} \right )~ e^{-ih\theta} d\theta \\
&=&\frac{1}{2\pi}\int_0^{2 \pi}\sum_{q,q'=1}^{\infty} \sum_{\stackrel{k=1}{(k,q)=1}}^q\sum_{\stackrel{k'=1}{(k',q')=1}}^{q'}
a_q a_{q'} \delta(\theta-2\pi (\kq-m) )\delta ( - \theta -2\pi(\kqp -n)) e^{-i h\theta} d\theta \\
&=&\sum_{q,q'=1}^{\infty} \sum_{\stackrel{k=1}{(k,q)=1}}^q\sum_{\stackrel{k'=1}{(k',q')=1}}^{q'}
a_q a_{q'} e^{-2\pi ih \kqp} \sum_{m,n= -\infty}^{\infty} \delta ( 2\pi(\kq -m -(\kqp -n))) e^{2\pi ihn}\\
&=& \sum_{q=1}^{\infty} a_q^2 c_q(h) \sum_{m=-\infty}^{\infty} 1
\end{eqnarray*}
We write this as
$$
\frac{{\di \sum_{n =-\infty}^{\infty}}a_n a_{n+h}}{{\di \sum_{m=-\infty}^{\infty}} 1} ~=~
\sum_{q=1}^{\infty} a_q^2 c_q(h) \, .
$$

This formula has ratio of two infinite quantities which we interpret as follows.
The denominator is the number of integers and the numerator is a suitable function
on the integers. Hence the ratio is in some sense a density. Hence this quantity
is the density of twin primes if we take $a(n) = \Lambda (n) {\di \frac{\phi (n)}{n}}$. We note that though we are dealing with the ratio
of two infinite quantities which is exactly like renormalisation in quantum
field theory the argument can be made rigorous by a limiting process which
should be akin to regularisation in physics. The authors have not been able to find
a neat and clever regularisation which would then make the argument both simple
and precise. However, it should be emphasized that absolutely no hard analysis
of error terms is necessary in this case. It would be important to get simple
regularisation schemes as large  number of problems in additive number theory
fall in this category. As this method makes error term analysis unnecessary,
sharpening this argument would mean real progress in this field.

\vspace{1cm}
\noindent {\bf Acknowledgements} 

The authors wish to thank Professor H.S. Sharatchandra
for constant encouragement. The second author wishes to thank CSIR for
financial support.

\vspace{1cm}

\noindent {\bf References}
\baselineskip=15pt
\begin{description}
\item{[1]} G.E. Andrews, q-series, their development and application
in Analysis, Number theory, Combinatorics, Physics and Computer Algebra,
66, Regional Conference series in Mathematics, 1985.

\item{[2]} R.D. Carmichael, Expansions of arithmetical functions in infinite series, Proc. London Math. Soc. (2) 34(1932) 1 - 26.

\item{[3]} D.V. Chudnovski and G.V. Chudnovski, Classical and Quantum models and Arithmetic Problems, Lecture Notes in Pure and Applied Mathematics,
Vol. 92.

\item{[4]} G.H. Gadiyar and R. Padma, Ramanujan - Fourier series, The
Wiener - Khintchine formula and the distribution of prime pairs,
Preprint.

\item{[5]} G.H. Hardy, Note on Ramanujan's trigonometrical function $c_q(n)$ and certain series of arithmetical functions, Proc. Camb. Phil. Soc. 20(1921), 
263 - 271.

\item{[6]} G.H. Hardy and J.E. Littlewood, Some problems of `Partitio Numerorum'; III: On the expression of a number as a sum of primes, 
Acta Math. 44 (1922), 3, 1-70.

\item{[7]} S. Ramanujan, On certain trigonometrical sums and their applications in the theory of numbers, Trans. Camb. Phil. Soc., 22(1918), 259 - 276.

\item{[8]} A. Terras, Harmonic Analysis in Symmetric Spaces and applications,
I, Springer - Verlag, Heidelberg, Berlin, New York, 1985.

\item{[9]} M. Waldschmidt et al (Ed.), From Number theory to physics,
Springer - Verlag, Berlin, Heidelberg, 1992.
\end{description}
\end{document}